\documentclass[twocolumn,aps,floatfix,showpacs,showkeys,epsfig,graphics]{revtex4-1}%
\usepackage{graphicx}
\usepackage{amsmath}
\usepackage{amsfonts}
\usepackage{amssymb}
\usepackage{filecontents}
\usepackage{subcaption}

\newcommand{\newc}{\newcommand}
\newc{\beq}    {\begin{equation}}
\newc{\eeq}    {\end{equation}}

\newc{\beqa}    {\begin{eqnarray}}
\newc{\eeqa}    {\end{eqnarray}}
\newc{\bs}    {\section}
\newc{\no}    {\\ \nonumber}

\def\apj{{\em Astrophys. J.  }}
\def\apjl{{\em Astrophys. J. Lett. }}

\newc{\st}    {\stackrel}

\begin{document}
\title{  Quantum Entanglement of  Dark Matter}
\author{Jae-Weon Lee}\email{scikid@jwu.ac.kr}
\affiliation{Department of Renewable Rnergy, Jungwon University,
            85 Munmu-ro, Goesan-eup, Goesan-gun, Chungcheongbuk-do,
              367-805, Korea,}
              \affiliation{Department of Physics, North Carolina State University, Raleigh, NC 27695, USA}

\begin{abstract}
We suggest that the dark matter in the universe  has quantum entanglement
if the dark matter is  a Bose-Einstein condensation of ultra-light scalar  particles.
In this theory, any two regions of a galaxy are quantum entangled due to the quantum nature of the condensate.
We calculate the entanglement entropy of a typical galactic halo, which turns out to be
at least $O(ln(M/m))$, where $M$ is the mass of the halo and $m$ is the
mass of a  dark matter particle.
 The entanglement can be inferred from the rotation curves of the galaxy or
  the interference patterns of the dark matter density.
\end{abstract}

\maketitle

\section{Introduction}
Dark matter (DM) is one of the
greatest mysteries in  physics and astronomy.
Despite decade-long  efforts, the origin of  DM defies a successful explanation.
The flatness of the  galactic
rotation curves implies the presence of invisible DM
in galactic halos ~\cite{1990PhRvL..64.1084P}, and any plausible DM theory should explain the curves.
Cold dark matter (CDM) with the cosmological constant ($\Lambda$CDM) model   is  very successful in explaining the
large-scale structure of the universe, but it
encounters many difficulties at  the galactic scale and below.
 For example, numerical simulations with the $\Lambda$CDM model predict a cusped central density
 and too
many subhalos compared to observations~\cite{Salucci:2002nc,navarro-1996-462,deblok-2002,crisis}.

Although several proposals consider dark subhalos or baryonic feedback mechanisms to alleviate these concerns,
the consideration of an alternative to  CDM
 playing the role of  CDM at the super-galactic scale
and at the same time  suppressing sub-galactic structures is desirable.
Bose-Einstein
condensate (BEC) DM or  scalar field dark matter (SFDM)  ~\cite{sin1,myhalo}
 may be a good alternative to  CDM. In this model,
 DM is  a BEC of  ultra-light scalar particles with mass $m\simeq 10^{-22} eV$ \cite{fuzzy}, which
implies a large DM Compton wavelength comparable to a galaxy core.
This long wavelength prevents the formation of DM-dominated structures smaller than a dwarf galaxy.
Beyond this scale, the coherent nature of  BEC DM makes it behave like  CDM,
 hence  solving the problems of  CDM.
This behavior was confirmed by a recent high precision numerical study~\cite{2014NatPh..10..496S,Schive:2014hza}.

The idea that DM is an ultra-light boson condensate has a long history.
(See Refs. \citealp{2009JKPS...54.2622L,2014ASSP...38..107S,2014MPLA...2930002R,2014PhRvD..89h4040H,2011PhRvD..84d3531C,2014IJMPA..2950074H}
 for a  review.)
Baldeschi {\it et al.}~\cite{1983PhLB..122..221B} considered galactic halos of
self-gravitating bosons, and
Membrado {\it et al.}~\cite{1989PhRvA..39.4207M} obtained
the rotation curves for a self-gravitating boson sphere.
Sin \cite{sin1}  suggested that galactic halos
were   like gigantic atoms made of
ultra-light BEC DM such as
pseudo Nambu-Goldstone bosons. In that model,
the boson DM particles are described by using a single macroscopic wave function of $kpc$ size,
 and the quantum uncertainty principle
prevents halos from self-gravitational collapse.
  In the context of field theory and general relativity, Lee and Koh ~\cite{myhalo}
  suggested that DM halos are  giant boson stars described by the Einstein-Klein-Gordon (EKG) Equations.
Widrow and Kaiser \cite{1993ApJ...416L..71W} suggested a new numerical method
utilizing the Schr\"{o}dinger equation  for collisionless matter.
Similar ideas have been  developed  by many authors
in terms of fuzzy DM, BEC DM, wave DM or ultra-light axions ~\cite {Schunck:1998nq, PhysRevLett.84.3037,PhysRevD.64.123528,PhysRevD.65.083514,repulsive,fuzzy,
 corePeebles,Nontopological,PhysRevD.62.103517,Alcubierre:2001ea,2012PhRvD..86h3535P,2009PhRvL.103k1301S,
 Fuchs:2004xe,Matos:2001ps,0264-9381-18-17-101,PhysRevD.63.125016,Julien,Boehmer:2007um, Matos:2001ps}.
These models of dark mater (Scalar field dark matter: SFDM hereafter) have been shown
to be able to explain the  observed rotation curves
~\cite{PhysRevD.64.123528,0264-9381-17-1-102,Mbelek:2004ff,PhysRevD.69.127502,Guzman:2015jba,Hui:2016ltb},
 the large scale structures of the universe~\cite{2014NatPh..10..496S}, the minimum mass and the size of galaxies ~\cite{Lee:2008ux,Lee:2015cos},
 the cosmic background radiation,
spiral arms~\cite{2010arXiv1004.4016B} and more.
 Works  have also been done on the  stability of BEC halos ~\cite{Guzman:2013coa}.

On the other hand, quantum
entanglement (a nonlocal quantum correlation)~\cite{nielsen}
 is  an important physical resource for  quantum information processing such as the quantum key distribution and
  quantum computing. Furthermore, entanglement is considered as a new order parameter for condensed matter
 systems such as a BEC.
 Recently, entanglement  was  proposed to be the key to understand dark energy \cite {myDE}, gravity \cite{Lee:2010fg} and
 even the spacetime itself~\cite{VanRaamsdonk:2010pw}.
Several other works link cosmology with entanglement. (See  Ref. \citenum{MartinMartinez:2012sg} for a review.)
 These works usually focuse on the generation of quantum field  entanglement
  during the cosmic expansion, especially during inflation ~\cite{Nambu:2008my}.
  Quantum fluctuation, and hence, quantum entanglement of the cosmic vacuum can  influence the density perturbation
  during and after the inflationary phase. From these perspectives, understanding the entanglement among  DM particles is meaningful.

 In this paper, we show that the SFDM in the universe has entanglement among DM particles. Although DM makes up about 26\% of the universe, its entanglement has not been studied so far.
  Because BEC shows a macroscopic quantum behavior by definition, an expectation that the SFDM would also have galactic-scale quantum entanglement is reasonable.
In section II, we study the entanglement of the SFDM in a dwarf galaxy.  In section III, we show how to decide the entanglement from observations.
In section IV, we discuss our findings.

\section{Entanglement of Bose-Einstein Condensate Dark Matter}
SFDM behaves as a coherent wave with a $kpc$ wavelength while
 conventional CDM is usually treated as classical point particles.
In the SFDM model \cite{sin1},  the galactic DM halo is a single giant boson star, which can be described with a wave function $\psi(\bold{r})$ of
the non-linear Schr\"{o}dinger equation (the Gross-Pitaevskii equation )
or equivalently with
the following  Schr\"{o}dinger-Poisson equations (SPEs):
\beqa
\label{sch}
i\hbar \partial_{{t}} {\psi} &=&-\frac{\hbar^2}{2m} \nabla^2 {\psi} +m{\Phi} {\psi}, \no
\nabla^2 {\Phi} &=&{4\pi G} m|\psi|^2
\eeqa
%
with a self-gravitation potential $\Phi$,
where $m$ is the mass of an individual DM particle.
Note that the number density of local dark matter  is proportional to $|\psi(r)|^2$.
 $\Phi$ plays the role of a trap potential for the galactic BEC.
For simplicity, we  ignore visible matter in this paper,
which is a good approximation for DM-dominated galaxies.
The SPEs can be obtained from the mean field approximation of a many-body BEC Hamiltonian
 or the Newtonian approximation
of the SFDM Lagrangian ~\cite{myhalo}.

For the study of the entanglement of the SFDM, the SPEs are  not enough, and we need a multi-particle formalism.
In the many-body theory of a BEC,
one usually decomposes the field operator with  the annihilation operator $a_\alpha$ as
\beq
\label{expansion}
\hat{\Psi}(\bold{r})=\sum_\alpha \psi_\alpha(\bold{r}) a_\alpha,
\eeq
where
the single-particle wave functions $\psi_\alpha(\bold{r})$ are orthonormal, i.e.,
\beq
\sum_\alpha \psi_\alpha(\bold{r}) \psi_\alpha^*(\bold{r'})=\delta(\bold{r}-\bold{r'}).
\eeq
Then,  $\hat\Psi(\bold{r})$ fulfills the commutation relation $[\hat{\Psi}(\bold{r}),\hat{\Psi}^\dag(\bold{r'})]=\delta(\bold{r}-\bold{r'})\mathbf{1}$.
The SFDM has a high phase-transition temperature ($T_c \gg TeV$), and the present temperature of  DM is so low that
we can treat a SFDM state as a zero-temperature BEC state.
At zero temperature, all boson particles are in the ground state, which is represented by
\beq
|\Psi_0\rangle=\frac{{a_0}^{\dag N}}{\sqrt{N!}} |0\rangle,
\eeq
where $N$ is the number of DM particles, $|0\rangle$ denotes the vacuum state, and
$a_0^\dag=\int d^3\bold{r} \psi_0 (\bold{r})\hat{\Psi}^\dag (\bold{r})$.
 $\psi_0 (\bold{r})$  is the lowest energy mode function for a single particle in a boson halo, which can be obtained by solving the SPEs (Eq. (\ref{sch})) .

Studies on the entanglement of a BEC in labs usually focuses on a spinor BEC or a multi-component BEC. However, for our purpose, we need
a formalism for a single-component scalar BEC.
Following the formalisms in Refs. \citealp{PhysRevA.66.052323} and ~\citealp{PhysRevA.80.012329}, we consider the entanglement between two subregions,  A and B, of a galaxy.
For example, the regions $A$ and $B$ can be the central part and an outer part of a galaxy, respectively.

For the calculation of  the entanglement between two regions of a galaxy,
  annihilation operators ($a_A,a_B$) for subsystems  are conveniently defined as
\beqa
a_A^\dag =\frac{1}{P_A}\int_{\bold{r}\in A} d^3 \bold{r}~ \psi_0(\bold{r}) \hat{\Psi}^\dag (\bold{r}), \no
a_B^\dag =\frac{1}{P_B}\int_{\bold{r}\in B} d^3 \bold{r}~\psi_0(\bold{r}) \hat{\Psi}^\dag (\bold{r}),
\eeqa
where $P_A=\int_{\bold{r}\in A} d^3 \bold{r} |\psi_0(\bold{r})|^2=1-P_B=
1-\int_{\bold{r}\in B} d^3 \bold{r} |\psi_0(\bold{r})|^2$.
 This means that $a_A^\dag$ creates a DM particle in  subregion A, and the probability to find the particle there is $P_A$.
Then, the ground state becomes
\beqa
\label{Psi0}
|\Psi_0\rangle &=&\frac{1}{\sqrt{N!}}(\sqrt{P_A}a^\dag_A + \sqrt{P_B} a^\dag_B)^N |0\rangle \no
&=& \frac{1}{\sqrt {N!}}\sum_{k=0}^N {{}_N C_k}  \sqrt{ P_A^{{k}} P_B^{{N-k}}}a^{\dag k}_A a^{\dag (N-k)}_B |0\rangle  \no
&=&  \sum_{k=0}^N\sqrt{ {{}_N C_k}  P_A^{{k}} P_B^{{N-k}}} |k\rangle_A|N-k\rangle_B \no
&\equiv& \sum_{k=0}^N \sqrt{\lambda_k} |k\rangle_A|N-k\rangle_B,
\eeqa
 where $a^{\dagger k}  |0\rangle =\sqrt{k!}  |k\rangle $ and  $\lambda_k={}_NC_k P_A^k P_B^{N-k}$.
Note that the final state is not separable; thus, $any$ part  of DM in a galactic halo is entangled with the DM in other regions in the same galaxy.

The measure of entanglement we choose is  the entanglement entropy
\beq
S_E=-\sum_k \lambda_k ln \lambda_k,
\eeq
which is the entropy of the reduced density matrix $\rho_A\equiv Tr_B(|\Psi_0\rangle \langle \Psi_0|$).
As $N\rightarrow \infty$,  the De Moivre-Laplace theorem
\beq
\lambda_k={}_NC_k P_A^k P_B^{N-k}\simeq \frac{1}{\sqrt{2\pi N P_A P_B}} exp\left(-\frac{(k-NP_A)^2}{2NP_A P_B}\right)
\eeq
implies that the reduced density matrix has a normal distribution with entropy
\beqa
\label{newSE}
S_E(\lambda_k)&\simeq& \frac{1}{2} ln(2 \pi e N_A P_B) \no
&=&\frac{1}{2} ln\left(2 \pi e \frac{M_A}{m}(1-\frac{M_A}{M})\right) \no
&=&\frac{1}{2} ln\left(\frac{M_A}{m}(1-\frac{M_A}{M})\right) +1.42,
\eeqa
where $N_A=NP_A$ is the mean particle number in subregion $A$ and $M_A=m N_A$ is the mass of the subregion A.
Note that  by definition $S_E$ is symmetric under  an interchange of $A$ and $B$.
This means   $S_E$ has a maximum value $\frac{1}{2} ln( \pi e N_A )$ when $P_A=P_B=1/2$.
%
%

For $m=10^{-22} eV$ \cite{fuzzy} and $M_A \ll M$,
\beq
\label{SEMA}
S_E(\lambda_k) \simeq \frac{1}{2}ln (M_A/M_\odot)+102.42
\eeq
 with the solar mass $M_\odot$.
For a typical galaxy, the halo mass is $M\simeq 10^{11} M_\odot$, so we expect $S_E=O(10^2)$.
This is not a big number, but  this value is just for the entanglement for
 a specific bisection of a galaxy.
Furthermore,  entanglement may exist between other modes we did not consider.
Therefore, $S_E$ in Eq. (\ref{SEMA}) should be treated as a lower bound for the entanglement of a galaxy.

To see how the entanglement scales,
we consider a bipartite system that consists of a spherical subregion (A)  within a radial distance  $r$
 from the center of galaxy and an outer region (B) beyond $r$.
 These regions could be parts of dwarf galaxies or
  a solitonic core and a surrounding cloud of BEC, as shown in Ref. \citealp{2014NatPh..10..496S}. We first need to know the wave function
  to study the entanglement.

The  DM mass enclosed inside  $r$  is given by
\beq
\label{MA}
M_A(r)= 4\pi m\int_0^r dr' r'^2 |\psi_0(r')|^2,
\eeq
where $\psi_0$ is the single-particle wave function for the halo with the normalization  $4\pi\int_0^\infty dr' r'^2 |\psi_0(r')|^2=N=M/m$, with $M$ being  the total halo mass.
 Because in the SFDM model, all DM particles are in a single wave function,  $\psi_0$ is enough for the current calculation.
We expect the dominant contribution of the entanglement $S_E$ to be concentrated on the boundary region between  region A and region B, as usual.
However, in this paper, we do not separately consider the boundary effect.

Using $M_A(r)$, we can obtain the entanglement between A and B as a function of $r$  from Eq. (\ref{SEMA}).
Then, the rotation velocity
\beq
V_{rot}(r)=\sqrt{\frac{G M_A(r)}{r}},
\eeq
which can be observed,
is directly related to the entanglement
\beq
\label{SE2}
S_E(r) \simeq \frac{1}{2}ln \left(\frac{rV_{rot}^2(r)}{GM_\odot}\right)+102.42;
\eeq
 hence, we can infer the DM entanglement by observing the rotation curves of the galaxy.

To be concrete, let us consider a dwarf galaxy.
Dwarf galaxies are the smallest DM-dominated objects and are ideal for DM study.
The ground state of the SPEs is known to  explain the rotation curves of dwarf galaxies well~\cite{PhysRevD.64.123528,0264-9381-17-1-102,Mbelek:2004ff,PhysRevD.69.127502,Guzman:2015jba}.
 The SPEs for a dwarf galaxy have an approximate analytic solution $\psi_0(r)=\psi_0(0) e^{-r^2/r_c^2}$ ~\cite{2011PhRvD..84d3531C} for the single-particle ground state,
  where $r_c$ is the  core radius of the DM halo.

\begin{figure}[htbp]
\includegraphics[width=0.4\textwidth]{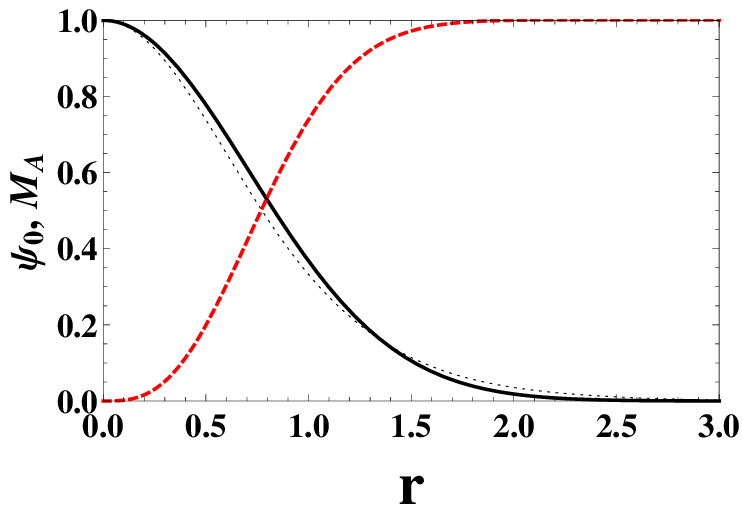}
\caption{(Color online) The approximate dark matter wave function $\psi_0$ (solid line) for a small galactic halo and
the rescaled mass inside $r$, $M_A(r)/M$ (red dashed line), as functions of
distance  ${r}$ (in units of the halo core radius $r_c$) from the halo center.
The dotted line represents the numerical solution of the SPE for  $\psi_0$.
The wave functions are rescaled as  $\psi_0(0)=1$.}
 \label{flat}
\end{figure}

\begin{figure}[htbp]
\includegraphics[width=0.4\textwidth]{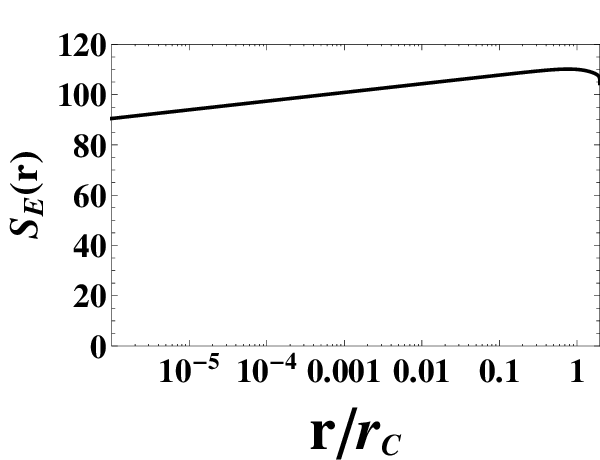}
\caption{ Entanglement entropy between a central spherical region and the outside of the region
as a function of  ${r}$ up to $3 r_c$.
}
 \label{flat}
\end{figure}

\begin{figure}[htbp]
\includegraphics[width=0.4\textwidth]{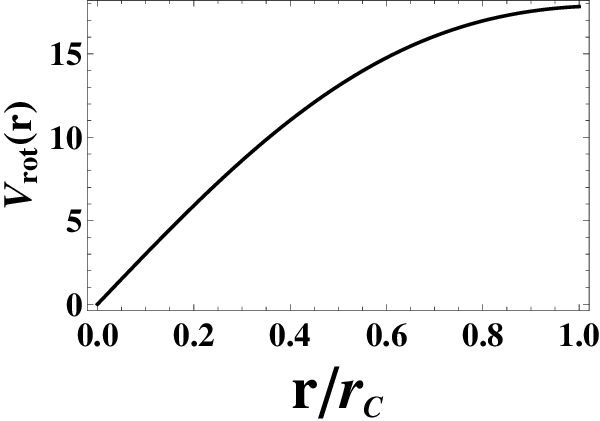}
\caption{ Rotation velocity in $km/s$ of the model halo with $r_c=100~pc$.
}
 \label{flat}
\end{figure}

  Figure 1 shows the above approximate solution,
   which looks quite similar to the numerical solution to the SPEs.
The approximate solution has no sharp boundary, and an arbitrary  a spatial cut off is chosen at $3 r_c$.
Then, the normalization condition $4\pi \int_0^{3r_c} dr r^2 |\psi_0(r)|^2=N$ gives $\psi_0(0)\simeq \alpha r_c^{-3/2}N^{1/2}$ with
$\alpha=0.713$.
Thus, the approximate ground-state wave function is
\beq
\psi_0(r)=\alpha N^{1/2}\frac{e^{-r^2/r_c^2}}{ r_c^{3/2}},
\eeq
and inserting Eq. (14) into Eq. (\ref{MA}) gives
\beq
M_A(r)= \frac{\pi\alpha^2}{4}\left(\sqrt{2 \pi } ~\text{erf}\left[\frac{\sqrt{2} r}{{r_c}}\right]
-4 e^{-\frac{2 r^2}{{r_c}^2}} \frac{r}{r_c}\right)M,
\eeq
where  erf denotes the error function. Now inserting $M_A(r)$ into Eq. (\ref{SEMA}) gives
 $S_E(r)$ as shown in Fig. 2.

Figure 1 and  2 show an example with $M=10^7 M_\odot$, which can be a model for a dwarf galaxy.
For comparison, the wave function is rescaled, and $M(r)$ is given in units of $M$.
Figure 2 shows $S_E(r)$, which is a rapidly rising function for  small $r$ and approaches a maximum value
as $M_A(r)$ approaches $M/2$.
One should  notice that our approximate formula  for $S_E$ is valid only for $ N_A, N_B \gg 1$;  hence,
it is
invalid  as    $r\rightarrow 0$ or as $r\rightarrow \infty$.
 However, obviously $S_E(r=0)=0$  because $|\Psi\rangle=|0\rangle_A|N\rangle_B$ if $N_A=0$.
$S_E$ is not small  even for a small $r$ because the particle mass $m$ is extremely small compared to  $M_A(r)$.
Even a tiny  region (say the solar system) has a gigantic number of DM particles that can be entangled with other regions.
Due to the long wavelength of the BEC DM particles, all parts of a galaxy are entangled with other parts of the galaxy.
Figure 3 shows the rotation curve for the model galaxy. Recall that $S_E(r)$ is related to $V_{rot}(r)$ via Eq. (\ref{SE2}).

\section{Entanglement and Interference}
How can we detect the  entanglement of  DM in the sky?
We can utilize the separability criterion based on interference
 for an atomic BEC in labs for the SFDM \cite{PhysRevA.66.052323, PhysRevA.68.062310}.
Let us consider three cases: 1)  typical CDM, 2) separable bosonic quantum DM  with a fixed $N$, and 3) entangled  bosonic quantum DM  with a fixed $N$ like the SFDM.
CDM particles such as  WIMPs (weakly interacting massive particles)  are usually treated as  classical point particles without coherence. Therefore, we cannot expect
any macroscopic quantum interference pattern or entanglement in galaxies made of typical CDM.

For separable  DM states, one may  expect some interference effects.
However,  the separable states cannot have perfect interference for a fixed $N$.
This can be shown by following the arguments for a BEC in Refs. 52 and 54.
All bipartite separable states have a density matrix in the form of
\beq
\rho=\sum_j p_j \rho^{(j)}_A \otimes\rho^{(j)}_B,
\eeq
where $\rho^{(j)}_A$ represents the $j$-th density matrix for the subsystem $A$.
Then,  according to the uncertainty principle, the separable states should satisfy the following inequality:
\beq
\label{inequality}
\left( (\Delta N)^2 +1 \right) \left( (\Delta(a_A-a_B))^2 +1\right) \ge \frac{\langle N\rangle_\rho} {4} +\frac{1}{8},
\eeq
where the real-valued
 variance for an observable $A$ is $(\Delta A)^2\equiv \langle A^\dagger A \rangle_\rho-| \langle A \rangle_\rho|^2$ for a given density matrix $\rho$
and $\langle N\rangle_\rho =N$ is the total number of DM particles.
This inequality is from the Cauchy-Schwarz inequality with the number operator $N=a^\dagger a.$ (See Eq. (4) of Ref. 54).
Thus, the separable states cannot have a very small particle number variation $\Delta N$
 $and$ perfect interference $(i.e., (\Delta(a_A-a_B))^2=0)$
at the same time for  large $N$.
Separable states with $(\Delta N)^2> O(N)$ can have clear interference  ~\cite{PhysRevA.68.062310},
 but  no  quantum DM model in this class has been found
  so far. Thus, we will not consider this exotic case  in this paper.
(The total mass or $N$ of  halo DM  is related to the gravitational influence of DM on astronomical objects around the halos.
   In a sense, this gives a continuous measurement of $N$ by the objects;   hence, we expect the DM state describing the halos as a whole
   to be one with  almost fixed $N$.)

On the other hand, entangled states can have very small particle number variation and clear interference at the same time,
violating the inequality in Eq. (\ref{inequality}) ~\cite{PhysRevA.68.062310}.
Therefore, the presence of a clear interference pattern in the DM distribution
 can be a plausible criterion for  DM entanglement.
This interference pattern in the halo's DM density distribution might be inferred from future precise observations of rotation curves, gravitational lensing~\cite{GonzalezMorales:2012uw}, and pulsa timing~\cite{Khmelnitsky:2013lxt,Aoki:2016mtn}, as well as from the data acquired by
gravitational wave detectors ~\cite{Aoki:2016kwl}.

As a toy example, we consider the DM interference fringes of two colliding galaxies $A$ and $B$ ~\cite{Gonzalez:2011yg}.
Note that now the regions $A$ and $B$ belong to two different colliding halos.
  If the SFDM particles in these galaxies are entangled, we would observe interferences violating the inequality in Eq. (\ref{inequality}).
  This situation is quite similar to the  interference experiment using an atomic BEC in a double-well potential in labs, so
  we can rely on the mathematics for an atomic BEC for  DM.
  By precisely observing the rotation curves or gravitational lensing, one can deduce the DM density distribution and $\langle N\rangle_\rho$,
  which is proportional to the total mass $M=m\langle N\rangle_\rho$.

\begin{figure}
\label{density}
\centering
 { \begin{subfigure}[t]{0.5\textwidth}
 \centering
    \includegraphics[width=\textwidth]{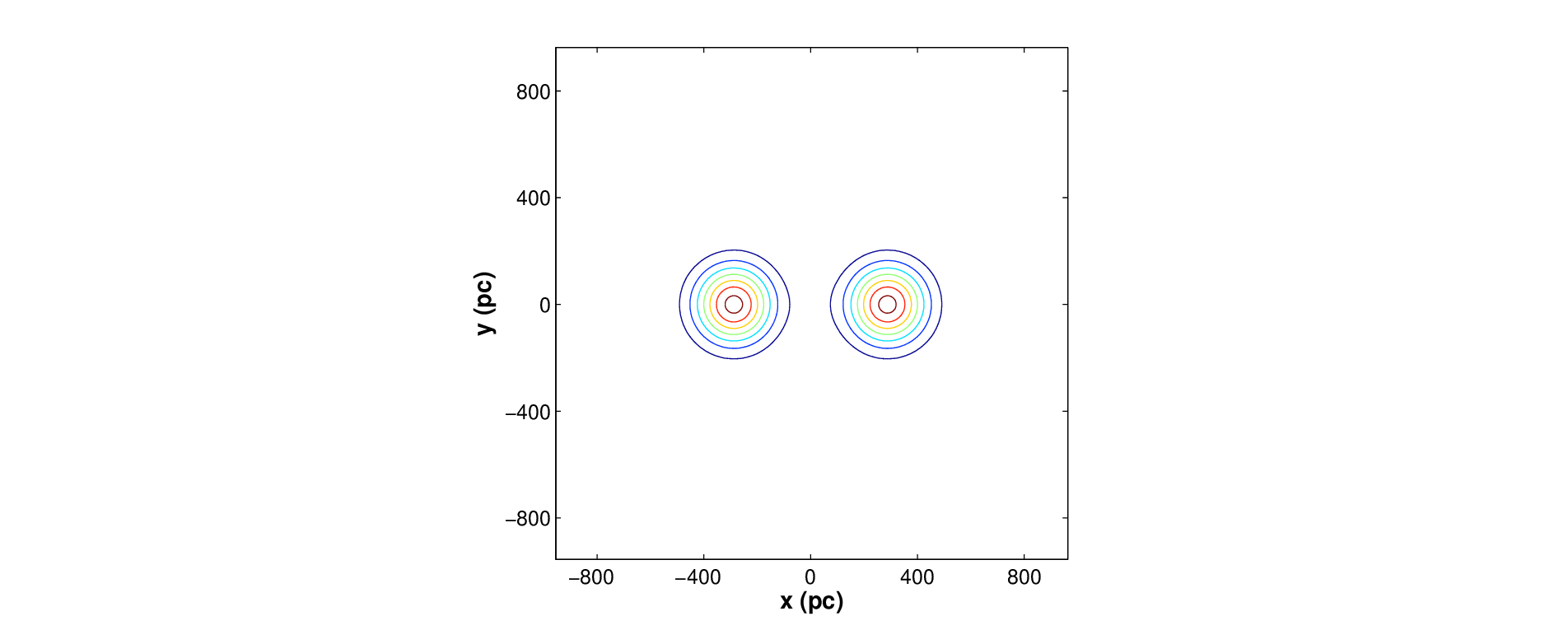}
    \caption{}
    \label{fig:1}
  \end{subfigure}
 } %
  \begin{subfigure}[t]{0.5\textwidth}
  \centering
    \includegraphics[width=\textwidth]{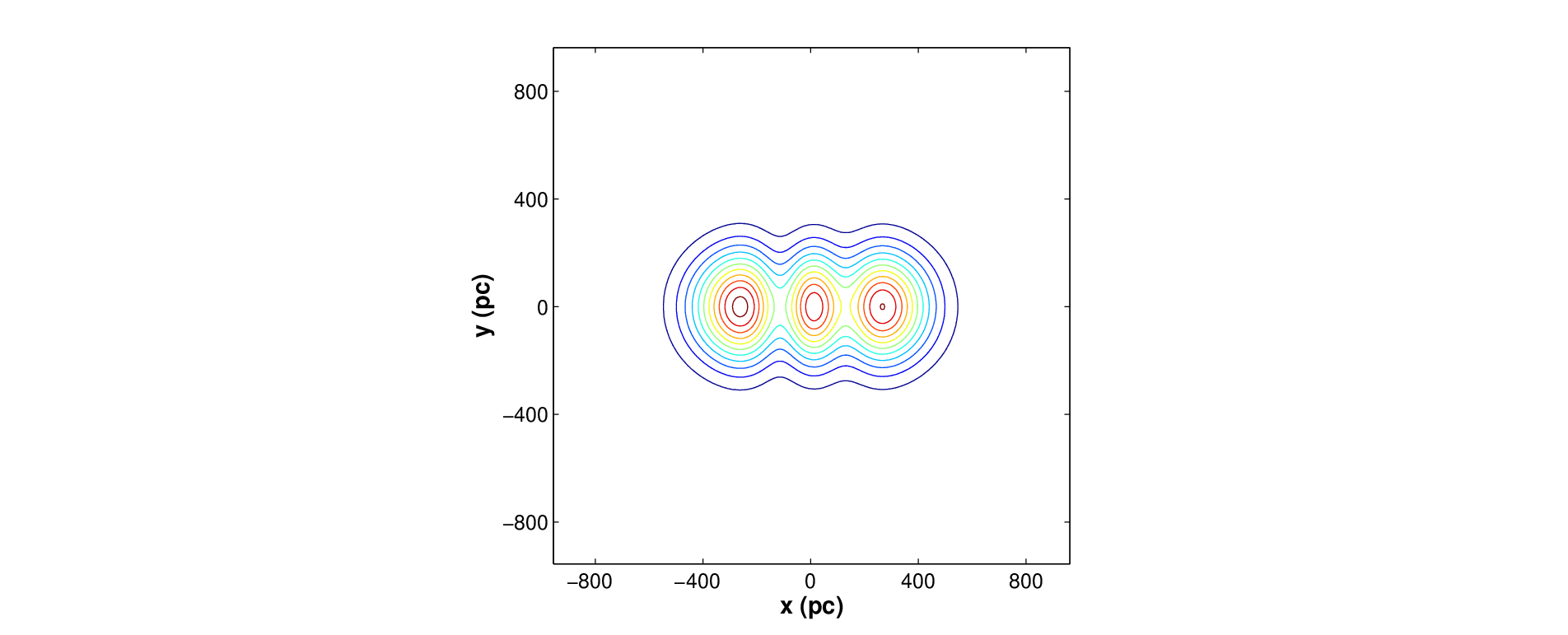}
    \caption{}
    \label{fig:2}
  \end{subfigure}
  \caption{A results of a numerical simulation showing the DM density distribution on the $x-y$ plane (a) before
    and (b) during  the collision of two equal galaxies with  total mass
   $M=10^7M_\odot$ and $\theta=0.2$. The interference fringe is clearly shown in (b) at $t=0.8~Myr$.}
\end{figure}

   The  density matrix $\rho=|\Psi_0\rangle \langle \Psi_0 |$ for the state in Eq. (\ref{Psi0}) of the SFDM has a  fixed $N$.
  Then, $(\Delta(a_A-a_B))^2$, which is related to the variance of the phase difference between
   modes $A$ and $B$ (i.e., two colliding DM halos),
  becomes a crucial quantity for the separability test.
  If the DM particles in each mode maintain the same phase, this variance should be exactly zero, which is impossible for  separable states
  with a fixed $N$.
  A  high-precision numerical study with the SFDM ~\cite{2014NatPh..10..496S} predicts
  the interference fringes made by the colliding DM halos. Therefore, we expect  the quantum entanglement to be universal in  SFDM-dominated cosmic structures.

  In  astronomical situations directly measuring the phase of the wave functions
  is difficult. However,
   because the variance $(\Delta(a_A-a_B))^2\le \langle (a^\dagger_A-a^\dagger_B) (a_A-a_B) \rangle_\rho\equiv  N_b $,  as a check of the inequality,
   measuring
  \beqa
  \label{Nb}
  N_b &\equiv& \langle b^\dagger b \rangle_\rho
  =\langle a^\dagger_A a_A + a^\dagger_B a_B
 -a^\dagger_A a_B -a^\dagger_B a_A \rangle_\rho  \no
 &=& N_A +N_B -\langle a^\dagger_A a_B +a^\dagger_B a_A \rangle_\rho,
  \eeqa
   which is the particle number in  mode $b\equiv a_A-a_B$, is sufficient
  ~\cite{PhysRevA.68.062310}.

 To understand the meaning of the last correlation term, we examine  two colliding galactic halos with a relative phase $\theta$ and a $50:50$ population ratio.
 The creation operators for this system  are
\beqa
\label{newa}
a_A^\dag =\frac{1}{\sqrt{2}}\int d^3 \bold{r}~ \psi_A(\bold{r},t) \hat{\Psi}^\dag (\bold{r}), \no
a_B^\dag =\frac{1}{\sqrt{2}}\int d^3 \bold{r}~\psi_B(\bold{r},t) \hat{\Psi}^\dag (\bold{r}),
\eeqa
where $\psi_A(\bold{r},t)\equiv u_A(\bold{r},t)e^{iQ_A \bold{r}}$ is the time-dependent single-particle wave function of the halo $A$ with
an initial  wave
vector $Q_A$;  similarly, $\psi_B(\bold{r},t)\equiv u_B(\bold{r},t)e^{iQ_B \bold{r}}$
 ~\citep{2007cond.mat..3766I}. $Q_A$ and $Q_B$ of the galaxies can be estimated from the physical properties of visible matter, such as the colliding velocity of hydrogen gas in each galaxy.
The $u_{A,B}$ is a real function with  normalization $\int d^3 \bold{r}~u_{A,B}^2=1$.
Then, the time-dependent entangled state similar to $|\Psi_0\rangle$  describing the pair of  halos is
\beq
|N,t\rangle\equiv \frac{1}{\sqrt{2^N N!}}\left(\int d^3\bold{r} \psi_A(\bold{r},t) e^{i \theta/2} + \psi_B(\bold{r},t) e^{-i \theta/2}\right)^N \hat{\Psi}^\dag (\bold{r}) |0\rangle,
\eeq
which is similar to the result for colliding two-BEC systems in the lab.
This time-dependent many-body quantum state represents the two colliding BEC DM halos ($A$ and $B$) with a relative global phase difference $\theta$
in their single-particle wave functions $\psi_A$ and $\psi_B$, respectably.
Now, we consider the Fourier component of the DM density distribution with a relative wave vector $Q\equiv Q_A-Q_B$;
\beqa
\label{rhoQ}
\langle \rho_Q\rangle &\equiv&\langle N,t|  \int d^3\bold{r} \hat{\Psi}^\dagger(\bold{r})  \hat{\Psi}(\bold{r}) e^{i Q \bold{r} } | N,t\rangle \no
&=&\langle N,t|  \int d^3\bold{r} (a_A^\dagger u_A e^{-iQ_A \bold{r}} +  a_B^\dagger u_B e^{-iQ_B \bold{r}}) \no
& &
(a_A u_A e^{iQ_A \bold{r}} +  a_B u_B e^{iQ_B \bold{r}}) e^{iQ\bold{r}}  | N,t\rangle \no
&=&\langle N,t|  a^\dagger_A a_B  | N,t\rangle \no
&=&\langle N-1,t| \frac{N e^{-i\theta}}{2} |N-1,t \rangle.
\eeqa
One can obtain the second equality by inverting  Eq. (\ref{newa}) and inserting $\hat{\Psi}(\bold{r})$.
The third equality came from the condition $Q=Q_A-Q_B$.
 After some straightforward calculations, one can obtain the last equality.
 (See Eq. (29) of Ref. \citealp{2007cond.mat..3766I} for details.)
We find $\langle \rho^\dagger_Q\rangle=\langle \rho_{-Q}\rangle=\langle N,t|  a^\dagger_B a_A  | N,t\rangle =Ne^{i\theta}/2$.
Therefore, for the entangled state $|N,t\rangle$,  $N_b$ in Eq. (\ref{Nb}) becomes
\beq
N_b= N_A + N_B -\langle N,t| \rho_Q +\rho^\dagger_Q  | N,t\rangle= N-N cos(\theta),
\eeq
which  violates the inequality (Eq. (\ref{inequality})) theoretically when $\theta < cos^{-1}(3/4)=0.723$  for  large $N$.
(This relation could also be useful for detecting the entanglement of an atomic BEC.)
In a sense, $N_b$ measures  a destructive interference. ($N_b$ is maximum for $\theta=\pi$.)

How can we observe $N_b$ astronomically in this case? One might obtain $\langle \rho_Q\rangle$ and, hence $N_b$ by carefully
 observing the DM density distribution
 \beq
 \rho(\bold{r})\equiv\langle N,t|  \hat{\Psi}^\dagger(\bold{r})  \hat{\Psi}(\bold{r})  | N,t\rangle
\eeq
in the colliding galaxies by observing rotation curves or gravitational lensing
 and taking the Fourier transformation of $\rho(\bold{r})$  by
 using Eq. (\ref{rhoQ}) with an estimated $Q$.
  Then, checking the deviation of the following inequality from Eq. (\ref{inequality})
  \beq
\label{inequality2}
N_b  \ge \frac{N} {4} -\frac{7}{8}
\eeq
is enough to decide the separability for the example we considered.
 In this way, the entanglement of galactic DM can be detected, in principle,
  by observing interferences, albeit  technically challenging.
Similarly, if we use   $\langle (a^\dagger_A+a^\dagger_B) (a_A+a_B) \rangle_\rho=N+N cos(\theta) $, we can detect a constructive interference.
As a byproduct, one can obtain the relative phase difference $\theta$ of two halos, too.

Figure 4 shows an example of an interference pattern obtained from a three-dimensional numerical solution of the SPEs.
We assumed  Gaussian wave packets for the two initial halos with $\theta=0.2$ and
an initial
relative velocity $4\times 10^{-4}c$.
Then, we
solved the SPEs by using the fast Fourier transformation code ~\cite{Paredes:2015wga}.
Separable DM such as  CDM cannot have  such a clear interference pattern as described above.

\section{Discussion}

We show that, in the SFDM theory,  two arbitrary regions of a galaxy are entangled even when  no self-interaction occurs among  DM particles
due to the nature of a BEC.
The entanglement between a region and other parts of the current observable universe can be similarly estimated using a formula.
A rough estimate gives $S_E\simeq \frac{1}{2} ln(M_u/M_\odot)+102.42\simeq 127.74$ at least,
where $M_u=O(10^{22}) M_\odot$ is the total DM mass of the universe.
The inflation period in the early universe
 might have provided a chance for quantum coherence among  widely separated cosmic structures;
however,  whether actual entanglement occurs among independent galactic halos is unclear.
The entanglement of  SFDM we consider is not from the inflationary phase, but from the generic property of a BEC.
On the other hand,  CDM particles such as WIMPs are usually very massive, and their number density is
too low to have a condensation. Thus, we expect  CDM particles not to have the macroscopic entanglement considered
in this paper.

Because SFDM  with an ultra-light mass  can be a viable alternative
to  CDM,  understanding the observational differences between the two models is important.
The SFDM can show a macroscopic quantum behavior because this DM has galactic-scale correlation and
coherence, and it
 is robust against
the decoherence induced by light or visible matter
 while conventional CDM particles are usually thought to be classical incoherent objects.
 A clear interference fringe in DM halos can appear in the SFDM model as a result of  entanglement.
Therefore,  entanglement of DM can be another criterion that can be used to
  distinguish  SFDM from  CDM,
for which observational effects, such as interference fringes, are  worth of more studies.

\acknowledgments
This work was supported by a Jungwon University Research Grant (2016-040).

\vskip 5.4mm

%

\end{document}